\documentclass[twocolumn,showpacs]{revtex4}

\usepackage{amsmath}
\usepackage{graphicx}
\usepackage{dcolumn}
\usepackage{epsfig}
\usepackage{verbatim}
\usepackage{psfrag}
\usepackage{bm}
\usepackage{bbm}
\usepackage{latexsym}
\usepackage[english]{babel}
\usepackage{amsthm}
\usepackage{color}
\usepackage{amssymb}

\begin{document}

\preprint{ITP-UU-11/05, SPIN-11/03}

\preprint{HD-THEP-11-2}

\title{Decoherence and Dynamical Entropy Generation in Quantum Field
Theory}

\author{Jurjen F. Koksma}
\email[]{J.F.Koksma@uu.nl} \affiliation{Institute for Theoretical
Physics (ITP) \& Spinoza Institute, Utrecht University, Postbus
80195, 3508 TD Utrecht, The Netherlands}

\author{Tomislav Prokopec} \email[]{T.Prokopec@uu.nl}
\affiliation{Institute for Theoretical Physics (ITP) \& Spinoza
Institute, Utrecht University, Postbus 80195, 3508 TD Utrecht, The
Netherlands}

\author{Michael G. Schmidt} \email[]{M.G.Schmidt@thphys.uni-heidelberg.de}
\affiliation{Institut f\"ur Theoretische Physik, Heidelberg
University, Philosophenweg 16, D-69120 Heidelberg, Germany}

\begin{abstract}
We formulate a novel approach to decoherence based on neglecting
observationally inaccessible correlators. We apply our formalism
to a renormalised interacting quantum field theoretical model.
Using out-of-equilibrium field theory techniques we show that the
Gaussian von Neumann entropy for a pure quantum state increases to
the interacting thermal entropy. This quantifies decoherence and
thus measures how classical our pure state has become. The
decoherence rate is equal to the single particle decay rate in our
model. We also compare our approach to existing approaches to
decoherence in a simple quantum mechanical model. We show that the
entropy following from the perturbative master equation suffers
from physically unacceptable secular growth.
\end{abstract}


\maketitle

\section{Introduction}

Decoherence is a quantum phenomenon that describes how a quantum
system turns into a classical stochastic system. The theory of
decoherence is widely used, e.g. in quantum
computing~\cite{QuantumComputing}, black hole
physics~\cite{Wald:1999vt}, inflationary perturbation
theory~\cite{Prokopec:1992ia} and in elementary particle physics
such as electroweak baryogenesis models~\cite{Farrar:1993sp}. It
is hard to study decoherence in quantum field theory (QFT) as it
requires out-of-equilibrium, finite temperature and interacting
quantum field theoretical computations. Let us first discuss the
conventional approach to decoherence~\cite{Zeh:1970} and its
shortcomings. We consider a system $S$ in interaction with an
environment $E$. The observer's inability to detect the
environmental degrees of freedom allows us to trace over $E$ to
construct a reduced density operator from the full density
operator \mbox{$\hat\rho_{\mathrm{red}} \!=\!{\rm Tr}_E[\hat
\rho]$}. The unitary von Neumann equation for the density operator
\mbox{$i
\partial_t \hat{\rho} \!=\! [\! \hat{H}\!,\!\hat{\rho}]$} is replaced by a
perturbative ``master equation'' for
\mbox{$\hat\rho_{\mathrm{red}}$} that is no longer unitary:
\begin{equation}\label{masterequation}
i \partial_t    \hat{\rho}_{\mathrm{red}} =
[\hat{H}_S,\hat{\rho}_{\mathrm{red}}]+
\mathcal{D}[\hat{\rho}_{\mathrm{red}}]\,,
\end{equation}
\noindent where \mbox{$\mathcal{D}[\hat{\rho}_{\mathrm{red}}]$}
collectively refers to all non-unitary dissipative terms
\cite{Zurek:2003zz, Caldeira:1982iu}. If
\mbox{$\hat{\rho}_{\mathrm{red}}$} is approximately diagonal in
the pointer basis, our quantum system effectively becomes a
classical stochastic system. In reality, however, systems do not
fully decohere so in order to quantify the amount of decoherence,
one is interested in the entropy increase due to the loss of
quantum coherence.

The master equation (\ref{masterequation}) suffers from both
theoretical and practical shortcomings. It is unsatisfactory that
\mbox{$\hat{\rho}_{\mathrm{red}}$} evolves non-unitarily while the
underlying QFT is unitary. On the practical side, the master
equation is so complex that basic field theoretical questions have
so far never been properly addressed: no well-established
treatment to take perturbative corrections to
\mbox{$\hat{\rho}_{\mathrm{red}}$} into account exists (for cases
in quantum mechanics (QM) see however \cite{Hu:1993vs}), nor has
any \mbox{$\hat{\rho}_{\mathrm{red}}$} ever been renormalised.
Moreover, as we will show, the perturbative master equation leads
even in very simple QM situations to physically unacceptable
secular growth of the entropy, which is caused by the perturbative
approximations used in specifying~(\ref{masterequation}).

The goal of this letter is twofold. Firstly, we present our novel
approach to decoherence that does not suffer from the shortcomings
mentioned above and apply it to a very simple QM toy model of
\mbox{$N\!+\!1$} coupled simple harmonic oscillators (SHO's).
Secondly, we apply our approach to an interacting QFT, using
non-equilibrium field theory techniques to calculate the time
evolution of the Gaussian von Neumann entropy. The focus and
novelty of this letter is not on calculational results, but
instead on a clear and succinct presentation of our approach to
decoherence and its relation to entropy generation in relativistic
quantum field theories. The results of various calculations we
discuss here serve mainly to test and illustrate our approach to
decoherence. We refer the reader for further details to
\cite{Koksma:2009wa, Koksma:2010zi, Koksma:2010dt, Koksma:2011dy}.

\section{Entropy and Correlators}

It is of course widely appreciated that entropy can be generated
as a result of an incomplete knowledge of a system. The infinite
hierarchy of irreducible $n$-point functions completely captures
all properties of a system, however only a finite subset can be
probed experimentally. This leads us to propose
\cite{Koksma:2009wa, Koksma:2010zi, Koksma:2010dt}: {\it
neglecting observationally inaccessible correlators will give rise
to an increase in entropy of the system as perceived by the
observer}. So both $S$ and $E$ evolve unitarily, however the
observer would say that $S$ evolves to a mixed state with positive
entropy as information about $S$ is dispersed in inaccessible
correlation functions. Giraud and Serreau \cite{Giraud:2009tn}
advocate similar ideas which they illustrate by an analysis in
\mbox{$\lambda \phi^4$,} building on earlier work
\cite{Prokopec:1992ia, Campo:2008ju}. The total von Neumann
entropy can be subdivided as follows:
\begin{equation}\label{vNeumanEntropySplit}
S_{\mathrm{vN}} = S^{\mathrm{g}}(t) + S^{\mathrm{ng}}(t) =
S^{\mathrm{g}}_S + S^{\mathrm{g}}_E + S^{\mathrm{g}}_{SE} +
S^{\mathrm{ng}}\,.
\end{equation}
\noindent Here, $S^{\mathrm{g}}$ is the total Gaussian von Neumann
entropy, that contains information about both $S$ and $E$ and the
their correlations $SE$, and $S^{\mathrm{ng}}$ is the total
non-Gaussian von Neumann entropy (consisting of $S$, $E$ and $SE$
contributions). In the following, we assume that the relevant
properties of quantum systems are encoded in the Gaussian part
\mbox{$\hat \rho_{\rm{g}}$} of the full density operator $\hat
\rho$. This is justified as higher order non-Gaussian $n$-point
functions are perturbatively suppressed. Our approach can be
improved if e.g. three- or four-point correlators are
experimentally accessible such that knowledge of these correlators
can be included in the definition of the entropy
\cite{Koksma:2010zi}. Although $S_{\mathrm{vN}}$ is conserved in
unitary theories, $S^{\mathrm{g}}_S(t)$ can increase at the
expense of other decreasing contributions to the total von Neumann
entropy, such as $S^{\mathrm{ng}}$. Calzetta and Hu
\cite{Calzetta:2003dk} prove an $H$-theorem for a quantum
mechanical \mbox{$O(N)$-model} and refer to ``correlation
entropy'' what we would call ``Gaussian von Neumann entropy''. For
e.g. the Gaussian system density matrix for a real scalar field,
one can straightforwardly find its associated von Neumann entropy
\cite{Campo:2008ju, Koksma:2010zi}:
\begin{equation}\label{Entropy_GaussianvNeumann}
S^{\mathrm{g}}_S \!= \! - \mathrm{Tr} [\hat{\rho}_{\mathrm{g}}\!
\ln \hat{\rho}_{\mathrm{g}}]\! =\!
\frac{\Delta\!+\!1}{2}\ln\!\Big[\frac{\Delta\!+\!1}{2}\Big] \!-\!
\frac{\Delta\!-\!1}{2}\ln\!\Big[\frac{\Delta\!-\!1}{2}\Big]\!.
\end{equation}
Here, $\Delta$ is the phase space area that the sytem's state
occupies in Wigner space ($\hbar\!=\!1$, $k\!=\!\|\vec{k}\|$):
\begin{equation}\label{PhaseSpaceArea1}
\Delta^2(k,t)  = 4 [\langle \hat \phi^2\rangle \langle \hat
\pi^2\rangle - \langle \{\hat \phi,\hat \pi\}/2\rangle^2 ] \,.
\end{equation}
\noindent The most efficient way of solving for the three Gaussian
correlators in (\ref{PhaseSpaceArea1}) is to solve for the
statistical propagator \mbox{$F_{\phi}(x;\!y)\! =\! \mathrm{Tr} [
\hat{\rho}(t_{0})\{ \hat{\phi}(x), \hat{\phi}(y) \}/2]$}, from
which the correlators can be straightforwardly extracted:
\begin{subequations}
\label{3 equal time correlators}
\begin{flalign}
\langle \hat{\phi}(\vec{x},t) \hat{\phi}(\vec{y},t) \rangle &=
F_{\phi}(\vec{x},t;\vec{y},t')|_{t=t'}\label{3 equal time
correlatorsa}
\\
\langle \hat{\pi}(\vec{x},t) \hat{\pi}(\vec{y},t) \rangle &=
\partial_{t} \partial_{t'} F_{\phi}(\vec{x},t;\vec{y},t')|_{t=t'} \label{3 equal time
correlatorsb}
\\
\langle \{ \hat{\phi}(\vec{x},t), \hat{\pi}(\vec{y},t) \}/2
\rangle &=
\partial_{t'} F_{\phi}(\vec{x},t;\vec{y},t')|_{t=t'} \label{3 equal time
correlatorsc}\,.
\end{flalign}
\end{subequations}
For a pure state we have \mbox{$\Delta\!=\!1$} and
\mbox{$S^{\mathrm{g}}_S\!=\!0$}, whereas for a mixed state
\mbox{$\Delta \!>\!1$} and \mbox{$S^{\mathrm{g}}_S\!>\!0$}. The
phase space area is conserved by the evolution
\mbox{$\mathrm{d}/\mathrm{d}t[ \Delta^2]\! =\! 0$} in free
theories. As \mbox{$S^{\mathrm{g}}_S$} is the only invariant
measure of the entropy of a Gaussian state
\cite{SohmaHolevoHirota:1999}, we argue that
\mbox{$S^{\mathrm{g}}_S$} should be taken as \emph{the}
quantitative measure for decoherence. This is to be contrasted
with most of the literature where different, non-invariant
measures are proposed \cite{Zurek:2003zz, Giraud:2009tn}.

Finally, let us connect the standard approach to ours. The
expectation values in (\ref{3 equal time correlators}) can in
principle be calculated from the reduced density matrix, consider
e.g. \mbox{$\langle \hat \phi^2 \rangle \!=\!{\rm
Tr}_{S+E}[\hat\rho\hat \phi^2]\! =\! {\rm
Tr}_S[\hat\rho_{\mathrm{red}} \hat \phi^2]$}. Suppose one has been
able to solve for $\hat{\rho}$ in \emph{full generality}, one
would then be interested in evaluating a ``reduced von Neumann
entropy'': \mbox{$S_{\mathrm{vN}}^{\mathrm{red}}(t) \!=\! -
\mathrm{Tr} [\hat{\rho}_{\mathrm{red}}(t) \log
\hat{\rho}_{\mathrm{red}}(t)]$}. We thus conclude that in the
Gaussian approach we also advocate, one would find
\mbox{$S_{\mathrm{vN}}^{\mathrm{red}}\!=\!S^{\mathrm{g}}_S$}. The
main point of this letter is that solving for a renormalised
$F_{\phi}$ in an interacting QFT is more tractable than solving
for a full $\hat{\rho}$, as we have recently witnessed a further
development of sophisticated techniques in out-of-equilibrium QFT
\cite{Berges:2004yj}.

\section{Quantum Mechanics}

Let us consider the well-known SHO model of $N\!+\!1$ coupled
oscillators:
\begin{equation}\label{LagrangianQM}
L= \frac12 {\dot x}^2 - \frac12 \omega^2_0
 x^2 + \sum_{n=1}^N\Big(\frac12 {\dot q}_n^2 - \frac12
\omega_n^2 q_n^2  - \lambda_n q_n  x \Big)\,,
\end{equation}
where $x$ plays the role of $S$, and the $\{q_n\}$ of $E$. This
model provides us with a level playing ground to compare the two
approaches to decoherence as QM toy models like these are free
from the drawbacks previously mentioned: we can actually solve
(\ref{masterequation}), and we do not have to worry about
renormalisation. Let us start by considering our approach.
Equation (\ref{vNeumanEntropySplit}) simplifies to
\mbox{$S_{\mathrm{vN}}\! =\! S^{\mathrm{g}}_S \!+\!
S^{\mathrm{g}}_E \!+\! S^{\mathrm{g}}_{SE}$} as
\mbox{$S^{\mathrm{ng}}\!=\!0$}. The role of the neglected
non-Gaussianities \mbox{$S^{\mathrm{ng}}$} in the QFT case is
played in the quadratic QM model by the correlation entropy
\mbox{$S^{\mathrm{g}}_{SE}$}.
\begin{figure}[t!]
\includegraphics[width=0.97\columnwidth]{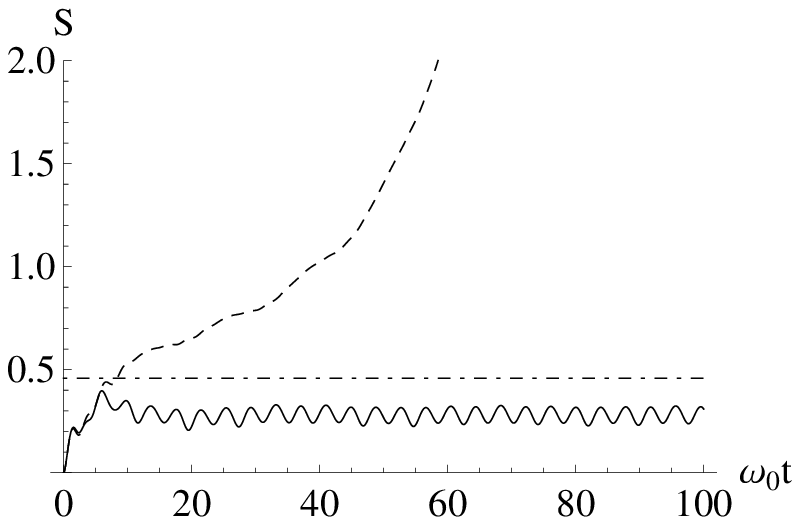}
\caption{Exact Gaussian von Neumann entropy $S^{\mathrm{g}}_S$
(solid) and the entropy from the master equation
(\ref{masterequation}) (dashed). We use \mbox{$\omega_n/\omega_0
\!=\! 1\!+\!n/100$,} \mbox{$1\! \leq \! n\! \leq \!50$,}
\mbox{$\lambda/\omega_0^2\!=\! 3/40$} and
\mbox{$\beta\omega_0\!=\!2$.} \label{fig:QMSHO}}
\end{figure}
Let us consider \mbox{$N\!=\!50$} oscillators, where all $E$
oscillators with frequencies $\{\omega_n\}$ act as a thermal bath
at temperature $\beta^{-1}$, \mbox{$k_{B}\!=\!1$.} Initially, we
require that $S$ is in a pure state
\mbox{$S^{\mathrm{g}}_S\!=\!0$}. Following our approach to
decoherence, we can numerically study the \mbox{\emph{exact}}
evolution of the three Gaussian QM analogues of equation (\ref{3
equal time correlators}): \mbox{$\langle \hat{x}^2\rangle$},
\mbox{$\langle \hat{p}_x^2\rangle$} and \mbox{$\langle
\{\hat{x},\hat{p}_x \}\rangle/2$.} In the resonant regime where
one or more \mbox{$\omega_n \! \sim \! \omega_0$,} the $\{q_n\}$
couple effectively to $x$ such that we expect to observe swift
thermalisation and decoherence. In figure \ref{fig:QMSHO} we show
the resulting evolution of the entropy (solid line). The
oscillations are a manifestation of Poincar\'e's recurrence
theorem \cite{Koksma:2010dt}. The dot-dashed line is the thermal
entropy corresponding to a temperature
\mbox{$\beta\omega_0\!=\!2$} indicating perfect thermalisation,
which at high temperatures, $\beta\omega_0 \! \ll \!1$, would
correspond to perfect decoherence. Given the weak coupling
\mbox{$\lambda/\omega_0^2\!=\!3/40$} we observe imperfect
decoherence (incomplete thermalisation). If we increase $\lambda$,
we observe full thermalisation of $S$ \cite{Koksma:2010dt}. We
conclude that a finite amount of entropy is generated at the
expense of obtaining a negative contribution to the correlation
entropy.

This behaviour is to be contrasted with the dashed line, the
master equation (\ref{masterequation}). The entropy resulting from
evolving $\hat{\rho}_{\mathrm{red}}$ suffers from physically
unacceptable unbounded secular growth. In cases where the entropy
obtained from the master equation settles to a constant value
before it breaks down, it generically overshoots. Let us stress
that we are in the deep perturbative (and resonant) regime and
that all eigenfrequencies following from (\ref{LagrangianQM}) are
positive. The perturbative master equation incorrectly resums
self-mass contributions, unlike the 2PI (two particle irreducible)
scheme available in QFT. Away from the resonant regime, where all
$\{\omega_n\}$ significantly differ from $\omega_0$, the evolution
of the entropy from the master equation agrees up to the expected
perturbative corrections with our approach, but this is the regime
where only a negligible amount of entropy is generated (no
decoherence). We conclude that decoherence rates at early times
can be calculated accurately using the standard approach to
decoherence however the perturbative master equation breaks down
at late times such that the total entropy increase, i.e.: the
total amount of decoherence, cannot be calculated reliably.

\section{Quantum Field Theory}

We will consider the following interacting scalar QFT:
\begin{equation}\label{actiontree1}
S[\phi,\chi] = \int \mathrm{d}^{\scriptscriptstyle{D}}\! x \{
{\cal L}_{0}[\phi] + {\cal L}_{0}[\chi]+ {\cal
L}_{\mathrm{int}}[\phi,\chi]\} \,,
\end{equation}
\noindent where:
\begin{subequations}
\label{actiontree2}
\begin{flalign}
{\cal L}_{0}[\phi] &= -\frac{1}{2} \partial_\mu\phi(x)
\partial_\nu \phi(x) \eta^{\mu\nu} - \frac{1}{2} m^{2}_{\phi}
\phi^{2}(x)
\label{actiontree2a}\\
{\cal L}_{\mathrm{int}}[\phi,\chi] &= -
\frac{\lambda}{3!}\chi^{3}(x) -\frac{1}{2}h \chi^{2}(x) \phi(x)
\,. \label{actiontree2c}
\end{flalign}
\end{subequations}
Here, $\phi$ will play the role of $S$ and $\chi$ of $E$. We
assume that \mbox{$\langle\hat{\phi}\rangle \!=\! 0\! =
\!\langle\hat{\chi}\rangle$}, which can be realised by suitably
renormalising the tadpoles. Equation (\ref{vNeumanEntropySplit})
reduces to \mbox{$S_{\mathrm{vN}} \!= \!S^{\mathrm{g}}_S \!+\!
S^{\mathrm{g}}_E \!+\! S^{\mathrm{ng}}$} as
\mbox{$S^{\mathrm{g}}_{SE}\!=\!0$}. The analogy with the simple QM
model is clear: in the QM case \mbox{$S^{\mathrm{g}}_S$} increases
as information is neglected in the Gaussian $SE$ correlations,
whereas in the QFT model \mbox{$S^{\mathrm{g}}_S$} increases as
non-Gaussian information is neglected, mainly in the $SE$
correlations. Note that in \cite{Giraud:2009tn} non-Gaussian $S$
correlations are neglected.

We work in the Schwinger-Keldysh or in-in formalism. The 2PI
effective action captures the effect of perturbative loop
corrections to the various propagators \mbox{$\imath
\Delta_{\phi}^{\! a b}$} and \mbox{$\imath \Delta_{\chi}^{\! a
b}$}, \mbox{$\{a,\!b\}\!=\!\pm$}. Varying the 2PI effective action
yields the Kadanoff-Baym (KB) equations for the Wightman
propagators:
\begin{equation}\label{KadanoffBaymEOM}
\!(\partial_{x}^{2}\!-\!m^{2}_{\phi})\imath \Delta^{\! \pm
\mp}_{\phi}\!(x;\! y)\! -\!\sum_{c=\pm}\!c\!\!\int\!
\mathrm{d}^{\!\scriptscriptstyle{D}}\! \tilde{x} M^{\pm
c}_{\phi}\!(x;\! \tilde{x})\imath \Delta^{\! c
\mp}_{\phi}\!(\tilde{x};\! y) \!=\!0 ,
\end{equation}
\noindent where the self-masses for $\phi$ at one loop read:
\begin{equation}
\imath M^{ab}_{\phi}(x;y) = -\frac{\imath h^{2}}{2} \left(\imath
\Delta^{ab}_{\chi}(x;y)\right)^{2} \,.
\end{equation}
\noindent Using dimensional regularisation, we find that a local
mass counterterm renormalises our theory \cite{Koksma:2009wa}:
\begin{equation}\label{SelfMasscounterterm}
\imath M_{\phi,\mathrm{ct}}^{\pm \pm}(x;y)= \mp \frac{\imath h^{2}
\Gamma(\frac{D}{2}-1) \mu^{\scriptscriptstyle{D}-4} }{16
\pi^{\frac{D}{2}} (D-3)(D-4)}\imath \delta^{\scriptscriptstyle{D}}
(x-y) \,.
\end{equation}
\begin{figure}[t!]
\includegraphics[width=0.97\columnwidth]{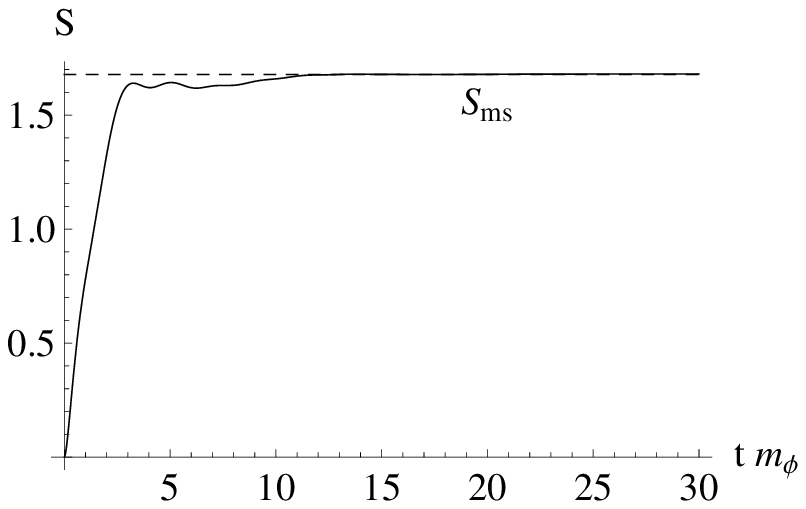}
\caption{$S^{\mathrm{g}}_S$ as a function of time (solid), which
quickly settles to \mbox{$S_{\mathrm{ms}}$} (dashed), where ``ms''
is an abbreviation for mixed state. We use \mbox{$\beta m_{\phi}
\!=\! 1/2$}, \mbox{$k/m_{\phi}\!=\!1$,} \mbox{$h/m_{\phi}\!=\!3$}
and \mbox{$N\!=\!2000$} up to \mbox{$t m_{\phi}\!=\!100$}.
\label{fig:SQFT}}
\end{figure}
\noindent The KB equations can be solved when written in terms of
the causal propagator \mbox{$\imath\Delta^{c}_{\phi} (x;\! y)\!
=\! \mathrm{Tr}( \hat{\rho}(t_{0}) [\hat\phi(x),\hat\phi(y)])\!=\!
\imath \Delta^{\! -+}_{\phi}\! -\! \imath \Delta^{\! +-}_{\phi}$}
and the statistical propagator \mbox{$F_{\phi}\! = \!
(\imath\Delta^{\! -+}_{\phi}\! +\! \imath\Delta^{\!
+-}_{\phi})/2$}. Initially, at \mbox{$t\!=\!t_0$,} $\phi$ is in a
pure state and $\chi$ is in thermal equilibrium at temperature
$\beta^{-1}$. We assume that \mbox{$\lambda\! \gg \!h$} such that
we can neglect the backreaction from $S$ on $E$. Perturbatively,
this argument is justified as the backreaction contributes only at
$\mathcal{O}(h^4\! /\omega_{\phi}^4)$:
\begin{equation}
\includegraphics[width=1.75cm]{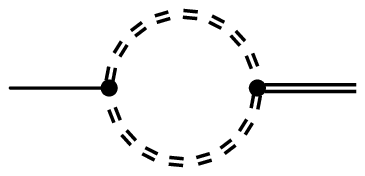}
\begin{tabular}{c}
\,\,=\,\, \\
\phantom{1} \\
\end{tabular}
\includegraphics[width=1.75cm]{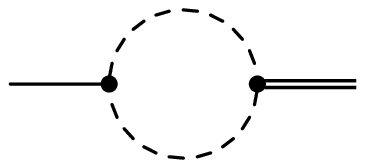}
\begin{tabular}{c}
\,\,+\,\, \\
\phantom{1} \\
\end{tabular}
\includegraphics[width=1.75cm]{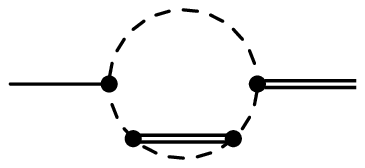}
\begin{tabular}{c}
\,\,+ \ldots\,\, \\
\phantom{1} \\
\end{tabular}
.\nonumber
\end{equation}
\noindent Here, a single (double) solid black line corresponds to
a free (resummed) $\phi$ propagator. Appreciate that the leading
order self-masses yield Gaussian corrections to \mbox{$\imath
\Delta_{\phi}^{\!ab}$}, so all non-Gaussianities are indeed
perturbatively suppressed. In our approximation scheme $\chi$
influences $\phi$ thus via the 1PI self-masses which are
calculated elsewhere \cite{Koksma:2009wa,Koksma:2011dy}. Memory
effects play an important role in non-equilibrium QFT and stem
from the time integrals
$\int_{-\infty}^{\infty}\!\mathrm{d}\tilde{t}$ over the
self-masses in (\ref{KadanoffBaymEOM}) with some cancellations
such that the final result is causal. We deal with the initial
time divergences by including memory effects at times
\mbox{$t\!<\!t_0$,} and an appropriate truncation method of these
integrals (for alternative approaches see \cite{Garny:2009ni},
e.g. Garny and M\"uller impose non-Gaussian initial conditions).

The resulting evolution for the entropy is shown in figure
\ref{fig:SQFT} which follows from $F_{\phi}$ and equations
(\ref{Entropy_GaussianvNeumann}--\ref{3 equal time correlators}).
Initially, we have  \mbox{$S^{\mathrm{g}}_S(t_0)\!=\! 0$} which
then rapidly increases and settles to its asymptotic value
\mbox{$S_{\mathrm{ms}}$}. We conclude that although a pure state
remains pure under unitary evolution, the observer perceives this
state over time as a mixed state with positive entropy
\mbox{$S_{\mathrm{ms}}$} as non-Gaussianities are dynamically
generated. In other words, a realistic observer cannot probe all
information about $S$ and thus discerns a loss of coherence of our
pure state.

\begin{figure}[t!]
\includegraphics[width=0.97\columnwidth]{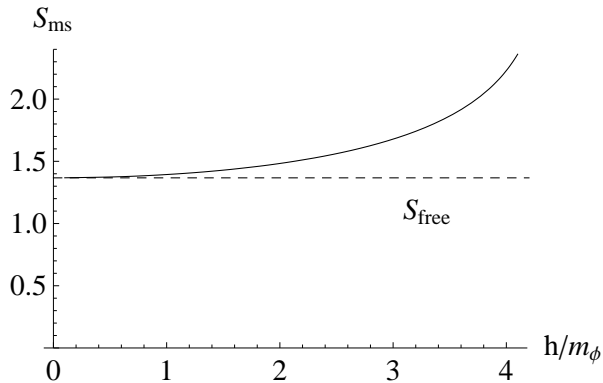}
\caption{The interacting thermal entropy $S_{\mathrm{ms}}$ as a
function of \mbox{$h/m_{\phi}$.} The dashed black line indicates
the free thermal entropy. We use \mbox{$\beta m_{\phi} \!=\! 1/2$}
and \mbox{$k/m_{\phi}\!=\!1$}. \label{fig:nonperturbativeregime}}
\end{figure}

We can extract two relevant quantities: the maximal amount of
decoherence and the decoherence rate. The maximal amount of
decoherence \mbox{$S_{\mathrm{ms}}$} can be calculated by
independent means and offers a powerful check of our approximation
scheme \cite{Koksma:2009wa, Koksma:2011dy}. At late times,
$F_{\phi}$ is time translationary invariant such that the entropy
is time independent, which can be appreciated from equation
(\ref{3 equal time correlators}), e.g. \mbox{$F_{\phi}(k,\!0)
\!=\! \int \! \mathrm{d}k^{0}\!/(\!2\pi\!) F_{\phi}(k^{\mu}\!)$}
is time independent. $F_{\phi}(k^{\mu}\!)$ follows directly from
the KB equations and the self-masses in Fourier space
\cite{Koksma:2009wa, Koksma:2011dy}. We show in figure
\ref{fig:nonperturbativeregime} that $S_{\mathrm{ms}}$ reduces to
the (free) thermal entropy when \mbox{$h\!\rightarrow \! 0$} given
by \mbox{$\Delta_{\mathrm{free}}\!=\!\coth(\beta\omega_{\phi}/2)$}
and equation (\ref{Entropy_GaussianvNeumann}). We thus see that
$S_{\mathrm{ms}}$ is the interacting thermal entropy.

\pagebreak

\begin{figure}[h!]
\includegraphics[width=0.97\columnwidth]{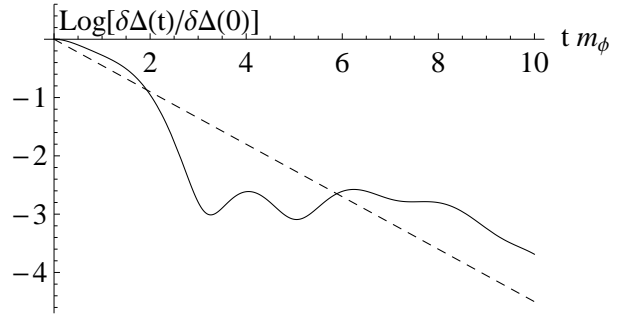}
\vskip 0.051cm \caption{Approach to $\Delta_{\mathrm{ms}}$ (solid)
and the decoherence rate $\Gamma_{\mathrm{dec}}$ (dashed). We use
the same parameters as in figure \ref{fig:SQFT}.
\label{fig:decoherencerate}}
\end{figure}

The decoherence rate is the rate at which the phase space area (or
entropy) changes. We observe as leading order effect an
exponential approach to $\Delta_{\mathrm{ms}}$ in figure
\ref{fig:decoherencerate}:
\begin{equation}\label{exponentialapproach}
\delta \Delta_{k}(t) \equiv \Delta_{\mathrm{ms}} - \Delta_{k}(t)
\simeq (\Delta_{\mathrm{ms}} - \Delta_{k}(0))
\exp[-\Gamma_{\mathrm{dec}} t] \nonumber \,.
\end{equation}

\noindent Here, $\Gamma_{\mathrm{dec}}$ is the decoherence rate of
our model which is well described by the single particle decay
rate of our interaction: \mbox{$ \Gamma_{\phi \rightarrow \chi
\chi} \!=\! - \mathrm{Im}(\imath
M^{\mathrm{r}}_{\phi})/\omega_{\phi}$}, where $\imath
M^{\mathrm{r}}_{\phi}$ is the retarded self-mass projected on the
quasi particle shell:
\begin{equation}
\Gamma_{\mathrm{dec}}\! \simeq \! \Gamma_{\phi \rightarrow \chi
\chi} \!=\! \frac{h^2}{32 \pi \omega_{\phi}} \!+\!  \frac{h^2}{16
\pi k \beta \omega_{\phi} } \log \!\left[
\frac{1\!-\!e^{\!-\frac{\beta}{2}(\omega_{\phi}+k)}}
{1\!-\!e^{\!-\frac{\beta}{2}(\omega_{\phi}-k)}}\right]\!.
\nonumber
\end{equation}

\section{Conclusion}

We have studied the time evolution of the Gaussian von Neumann
entropy in an interacting, out-of-equilibrium, finite temperature
QFT in a renormalised and perturbative 2PI scheme. We have
extracted two relevant quantitative measures of decoherence: the
maximal amount of decoherence and the decoherence rate. This study
builds the QFT framework for other decoherence studies in relevant
situations where different types of fields and interactions are
involved. In cosmology for example, the decoherence of scalar
gravitational perturbations can be induced by e.g. fluctuating
tensor modes (gravitons) \cite{Prokopec:1992ia}, isocurvature
modes \cite{Prokopec:2006fc} or even gauge fields. In quantum
information physics it is very likely that future quantum
computers will involve coherent light beams that interact with
other parts of the quantum computer as well as with an
environment~\cite{QuantumComputing}. For a complete understanding
of decoherence in such complex systems it is clear that a QFT
framework such as developed here is necessary.


\begin{thebibliography}{99}
\bibitem{QuantumComputing}
  M.~A.~Nielsen, I.~L.~Chuang,
  Cambridge University Press (2000);
  E.~Knill, R.~Laflamme, and G.~J.~Milburn,
  Nature {\bf 409} (2001) 46.
\bibitem{Wald:1999vt}
  R.~M.~Wald,
  Living Rev.\ Rel.\  {\bf 4 } (2001)  6.
\bibitem{Prokopec:1992ia}
  T.~Prokopec,
  Class.\ Quant.\ Grav.\  {\bf 10} (1993) 2295;
  R.~H.~Brandenberger, T.~Prokopec and V.~F.~Mukhanov,
  Phys.\ Rev.\  D {\bf 48} (1993) 2443;
  C.~Kiefer, I.~Lohmar, D.~Polarski and A.~Starobinsky,
  Class.\ Quant.\ Grav.\  {\bf 24 } (2007)  1699-1718.
\bibitem{Farrar:1993sp}
  G.~R.~Farrar, M.~E.~Shaposhnikov,
  Phys.\ Rev.\ Lett.\  {\bf 70 } (1993)  2833-2836.
\bibitem{Zeh:1970}
  H.~D.~Zeh,
  Found.\ Phys.\ {\bf 1} (1970);
  E.~Joos, H.~D.~Zeh, C.~Kiefer, D.~Giulini, J.~Kupsch and I.~O.~Stamatescu,
  Springer (2003).
\bibitem{Zurek:2003zz}
  W.~H.~Zurek,
  Rev.\ Mod.\ Phys.\  {\bf 75} (2003) 715.
\bibitem{Caldeira:1982iu}
  A.~O.~Caldeira, A.~J.~Leggett,
  Physica {\bf 121A } (1983)  587-616.
\bibitem{Hu:1993vs}
  B.~L.~Hu, J.~P.~Paz, Y.~Zhang,
  Phys.\ Rev.\  {\bf D47 } (1993)  1576-1594;
  B.~L.~Hu, A.~Matacz,
  Phys.\ Rev.\  {\bf D49 } (1994)  6612-6635.
\bibitem{Koksma:2009wa}
  J.~F.~Koksma, T.~Prokopec, M.~G.~Schmidt,
  Phys.\ Rev.\  {\bf D81 } (2010)  065030.
\bibitem{Koksma:2010zi}
  J.~F.~Koksma, T.~Prokopec, M.~G.~Schmidt,
  Annals Phys.\  {\bf 325 } (2010)  1277-1303.
\bibitem{Koksma:2010dt}
  J.~F.~Koksma, T.~Prokopec and M.~G.~Schmidt,
  Annals Phys.\  {\bf 326} (2011) 1548.
\bibitem{Koksma:2011dy}
  J.~F.~Koksma, T.~Prokopec and M.~G.~Schmidt,
  Phys.\ Rev.\ D {\bf 83} (2011) 085011.
\bibitem{Giraud:2009tn}
  A.~Giraud, J.~Serreau,
  Phys.\ Rev.\ Lett.\  {\bf 104 } (2010)  230405.
\bibitem{Campo:2008ju}
  D.~Campo, R.~Parentani,
  Phys.\ Rev.\  {\bf D78 } (2008)  065044;
  Phys.\ Rev.\  D {\bf 78} (2008) 065045.
\bibitem{Calzetta:2003dk}
  E.~A.~Calzetta, B.~L.~Hu,
  Phys.\ Rev.\  {\bf D68 } (2003)  065027.
\bibitem{SohmaHolevoHirota:1999}
  M.~Sohma, A.~S.~Holevo and O.~Hirota,
  Phys.\ Rev.\ A {\bf 59} (1999) 1820.
\bibitem{Berges:2004yj}
  J.~Berges,
  AIP Conf.\ Proc.\  {\bf 739} (2005) 3.
\bibitem{Garny:2009ni}
  M.~Garny, M.~M.~Muller,
  Phys.\ Rev.\  {\bf D80 } (2009)  085011;
  S.~Borsanyi and U.~Reinosa,
  Renormalised Nonequilibrium Quantum Field Theory: Scalar Fields,
  Phys.\ Rev.\  D {\bf 80} (2009) 125029.
\bibitem{Prokopec:2006fc}
  T.~Prokopec, G.~I.~Rigopoulos,
  JCAP {\bf 0711 } (2007)  029.
\end{thebibliography}
\end{document}